\documentclass[doublecol]
{epl2} 
\usepackage[utf8]{inputenc}

\usepackage{graphicx}
\usepackage{amssymb,amsfonts,amsmath}
\usepackage{color}
\usepackage{ulem}

\newcommand{\F}{\boldsymbol{F}}

\newcommand{\pvec}{\boldsymbol{p}}
\newcommand{\rr}{\boldsymbol{r}}
\newcommand{\nab}{\boldsymbol{\nabla}}
\newcommand{\xxi}{\boldsymbol{\xi}}
\newcommand{\xchi}{\boldsymbol{\chi}}
\newcommand{\eeta}{\boldsymbol{\eta}}
\newcommand{\R}{\boldsymbol{R}}

\begin{document}

\title{Active Brownian particles at interfaces: 
\\An effective equilibrium approach} 
\shorttitle{Active Brownian particles at interfaces}

\author{Ren\'e Wittmann \and Joseph M.~Brader} \shortauthor{Ren\'e Wittmann and Joseph M.~Brader}

\institute{Department of Physics, University of Fribourg, CH-1700 Fribourg, Switzerland}

\date{\today}

\abstract{
A simple theoretical approach is used to investigate active colloids at the free interface and near 
repulsive substrates. 
We employ dynamical
density functional theory to determine the steady-state density profiles in an
effective equilibrium system 
[Farage {\it et~al.}, \textit{Phys.\ Rev.\ E}, {\bf 91} (2015) 042310]. 
In addition to the known accumulation at surfaces, 
we predict wetting and drying transitions at a flat repulsive wall and capillary condensation and evaporation in a slit pore.
These new phenomena are closely related to the motility-induced phase separation (MIPS) in the bulk.
}


\maketitle

\section{Introduction}

Understanding self-organisation and non-equilibrium phase behaviour in active systems is currently a 
subject of intense research activity \cite{ramaswamy2010,cates_tailleur2014,swim_aniso}. 
Bulk systems of spherically symmetric, repulsive active Brownian particles (ABPs) have been demonstrated to 
undergo a motility-induced phase separation (MIPS) into coexisting high- and low-density 
non-equilibrium phases \cite{cates_tailleur2014,fily2012,cates2013,stenhammar2014,bialke2013,speck2014,buttinoni2013}. 
The phenomenon of MIPS is a consequence of the persistent trajectories of active particles; 
the particles run into each other and thus tend to cluster \cite{buttinoni2013}. 
Systems for which the passive interaction has an attractive component exhibit an even richer 
collective behaviour. 
In this case, it has been found that increasing activity can lead first to a suppression of passive phase 
separation \cite{schwarz-linek2012,redner2013,faragebrader2015}, followed by a re-entrant 
phase separation at high activity \cite{redner2013,faragebrader2015}.

A convenient strategy to describe the phenomenology of {active systems}
involves an {\it effective} equilibrium picture \cite{faragebrader2015,maggi,marconi}. 
The theory proposed in ref.~\cite{faragebrader2015} \mbox{accounts for} certain aspects of MIPS
for both repulsive and attractive interactions.
Within this approach {the non-Gaussian noise due to orientational fluctuations in the 
microscopic equations of motion of ABPs is replaced with appropriate Ornstein Uhlenbeck processes (OUPs).
Deriving an} approximate Fokker-Planck equation {for this model and taking the low-density limit
 yields} an effective pair potential, extending the 
bare interaction potential via parameters specifying the activity~\cite{faragebrader2015}.
The effective potential may be used as input to established liquid-state theories~\cite{bob79,bob92,hansen_mcdonald1986} to make 
predictions about phase equilibria and the microstructure
of particles propelled by OUPs and, for systems not too far from equilibrium~\cite{faragebrader2015}, also of ABPs.

Compared to the bulk, much less is known about active systems at interfaces,
although swimming microorganisms are naturally found in confinement, leading to surface accumulation
\cite{swim_aniso,kantsler2013}.
In some limiting cases the steady-state problem is tractable analytically on the single-particle level~\cite{enculescu2011,szamel2014,pototsky2012,elgeti2013},
 providing intuition for interacting active particles 
under gravity~\cite{ginot2015}, in a harmonic trap~\cite{pototsky2012} or running into substrates~\cite{xiao2014,yang2014,ni2015}.
Recent simulations of two-dimensional ABPs indicate more general interfacial phase transitions~\cite{ni2015}
and show that, despite a negative interfacial tension arising from interparticle forces, there exists a stable free interface between
coexisting MIPS states~\cite{speck_interface2014}.
A point of particular interest is now whether inhomogeneous steady states of ABPs
can be adequately described using an effective approach, which would enable experience 
from equilibrium thermodynamics to be exploited.   

In this Letter we combine the ideas of refs.~\cite{faragebrader2015} and \cite{pototsky2012}
to construct a general dynamical density functional theory (DDFT) \cite{marconi1999,archer_evans_spinodal2004} for inhomogeneous situations. 
The resulting effective equilibrium theory is used to
investigate the collective behaviour of a model for 
ABPs at the free interface, near a repulsive 
planar substrate and under confinement between two repulsive walls. 
Considering statepoints close to the MIPS phase boundary (binodal) we predict the phenomena 
of activity-induced wetting and capillary condensation in systems of repulsive ABPs. 
For attractive ABPs we observe analogues of drying and capillary evaporation. 
Finally, we discuss our approach regarding recent opinions about active thermodynamics~\cite{cates_tailleur2014,takatori2014,takatori2015,solon_BrownianPressure2015,speck_interface2014,cates_kardar,solonEPJST,catesarxiv,speckarxiv}.

\section{Theory}
We consider a system of interacting, spherical ABPs with coordinate $\rr_i$ and a self-propulsion
 of speed $v_0$ acting in the direction of orientation, specified by the unit vector 
$\pvec_i$.
The Langevin equations
\begin{eqnarray}
\dot{\rr}_i = v_0\,\pvec_i  + \smash{\gamma^{-1}\F_i^{\rm tot}} + \xxi_i\,,\ \ \  
\label{full_langevin}
\dot{\pvec}_i = \eeta_i\times\pvec_i\,
\end{eqnarray}
with the friction coefficient $\gamma$ and the total force $\F_i^{\rm tot}(\rr^N\!\!,t)$ on particle $i$
describe the motion of ABPs. 
The stochastic vectors $\xxi_i(t)$ and $\eeta_i(t)$ are Gaussian distributed with zero mean and 
have the time correlations	
$\langle\xxi_i(t)\xxi_j(t')\rangle\!=\!2D_\text{t}\boldsymbol{1}\delta_{ij}\delta(t-t')$ and 
$\langle\eeta_i(t)\eeta_j(t')\rangle\!=\!2D_\text{r}\boldsymbol{1}\delta_{ij}\delta(t-t')$, where 
$D_\text{t}\!=\!k_\text{B}T/\gamma\!=\!(\beta\gamma)^{-1}$ and $D_\text{r}$ are the translational and rotational diffusion coefficients, respectively, and $\beta$ denotes the inverse of the temperature $T$ with the Boltzmann constant $k_\text{B}$. 
The approximation methods adopted in the following to eliminate the orientational degrees of freedom are detailed in refs.~\cite{faragebrader2015,pototsky2012}.

\subsection{Effective interaction potential} 

{Representing the orientational degrees of freedom of ABPs by OUPs $\xchi_i$ with the same coloured-noise statistics 
as $v_0\,\pvec_i$ in eq.~\eqref{full_langevin}, i.e., 
$\langle\xchi_i(t)\xchi_j(t')\rangle\!=\!v_0^2\boldsymbol{1}\delta_{ij}\exp(-2D_r|t\!-\!t'|)/3$,
an} approximate Fokker-Planck equation for the configurational probability distribution $\bar{P}(\rr^N\!\!,t)$
{and thus} the following effective, activity-dependent pair potential can be identified~\cite{faragebrader2015} 
\begin{eqnarray}\label{potential_pair}
\beta u^{\text{eff}}(r)=-\int_r^\infty\!\mathrm{d}r'\left[\frac{\partial_{r'}\beta u(r')}{\mathcal{D}(r')}+\partial_{r'}\ln\mathcal{D}(r')\right]\,,
\end{eqnarray}
where $u(r)$ is the bare pair potential, 
$\mathcal{D}(r)\!=\!1\!+\!Pe^2\tau/(3\!+\!3\tau (r{/d})^{-2}\partial_r[r^2 \partial_r\beta u(r)])$ 
and $\partial_r\!=\!\partial/\partial r${, with the particle diameter $d$ entering $u(r)$.
The only} parameters, {describing the strength of activity}, are the dimensionless persistence 
time \mbox{$\tau\!=\!D_\text{t}/{(}2D_\text{r}d^{\,2}{)}$} and the P\'eclet number $Pe\!=\!v_0d/D_\text{t}$.

The position of the minimum of $u^{\text{eff}}(r)$ (if present) moves to shorter separations with increasing $Pe$ 
(see figs.~\ref{fig_plots}b and \ref{fig_LJ}a). 
This is due to the fact that active particles run into each other and spend more time 
at closer separations than in the corresponding equilibrium system. 
Within our picture this `self-trapping' effect is mimicked by lending such configurations 
additional statistical weight. 
In ref.~\cite{faragebrader2015} the effective potential was input to liquid-state integral-equation 
theory and found to accurately reproduce the phenomenology of MIPS in a system of ABPs.
We next extend the effective equilibrium approach to treat inhomogeneous problems.

\subsection{Effective external potential} 
From eq.~\eqref{full_langevin} it can be exactly shown that the 
joint probability distribution, $\!P(\rr^N\!\!,\pvec^N\!\!,t)$, of ABPs evolves according to 
\begin{eqnarray}\label{smol_eq2}
\!\partial_t P\!=\!\sum_{i=1}^{N} \nab_{i}\!\cdot\!
\big[
D_\text{t}\!\left(\nab_{i}\! - \beta\F_i^{\rm tot}\right)
\!-\!v_0\,\pvec_i\big] P \!+\! D_\text{r}\R_i^2 P, 
\end{eqnarray}
where $\R\!=\!\pvec\times\!\nab_{\!\pvec}$ is the rotation operator. 
Integration of eq.~\eqref{smol_eq2} over orientations and $N\!-\!1$ coordinates yields 
\begin{eqnarray}\label{DDFT_exact}
\partial_t \rho= D_\text{t}\,\nab\cdot\left(
\nab  - \beta\F^{\rm int}+ \nab \beta V_{\rm ext} -\beta\F^{\rm act}\right)\rho
\end{eqnarray}
This coarse-grained equation of motion 
for the one-body density $\rho(\rr,t)\!=\!\int \!\mathrm{d}\pvec\, \bar{\rho}(\rr,\pvec,t)$
involves the activity force 
\begin{eqnarray}\label{average_orientation}
\beta\F^{\rm act}(\rr,t) =D_\text{t}^{-1} v_0\!\int \mathrm{d}\pvec \,\,\pvec\,\frac{\bar{\rho}(\rr,\pvec,t)}{\rho(\rr,t)}
\end{eqnarray}
and the \textit{average} interaction force \vspace*{0.21cm}
\begin{eqnarray}\label{average_interaction}
\!\!
\F^{\rm int}(\rr,t)\!=\! -\!\!\smash{\iiint\! \mathrm{d}\pvec \,\mathrm{d}\pvec'\mathrm{d}\rr' \,\frac{\rho^{(2)}(\rr,\rr'\!,\pvec,\pvec'\!,t)\,
\nab u(|\rr\!-\!\rr'|)}{\rho(\rr,t)}}\,,
\notag
\hspace*{-1cm}
\\
\end{eqnarray}\vspace*{-0.5cm}

\noindent
where $\rho^{(2)}(\rr,\rr'\!,\pvec,\pvec'\!,t)$ is the non-equilibrium two-body density. 
To proceed we will approximate eqs.~\eqref{average_orientation} and 
\eqref{average_interaction}. 

We first address the activity force \eqref{average_orientation}, which can be approximated using the 
methods of Pototsky and Stark \cite{pototsky2012}.  
We specialise here to a planar geometry for which the density varies only 
in the $z$-direction. 
In the low-density limit the probability $P$ in eq.~\eqref{smol_eq2} factorises and 
the problem reduces to that of a single particle in a planar external field $V_{\rm ext}(z)$. 
Asymptotic analysis 
in terms of the small parameter $\tau$~\cite{pototsky2012} yields an approximate steady-state solution 
for the activity force on a single particle
\begin{eqnarray}\label{onebody_active}
\F^{\rm act}_{\rm ss}(z) = Pe
 \frac{Pe\tau\, \partial_z V_{\rm ext}(z)}{1 + Pe^2\tau\beta V_{\rm ext}(z)}. 
\end{eqnarray}
Equation \eqref{onebody_active} is easily integrated to obtain an 
activity-dependent potential. 
Together with the bare external field this yields the {\it effective} 
external potential 
\begin{eqnarray}\label{effective_external}
V^{\rm eff}_{\rm ext}(z) = V_{\rm ext}(z) - \beta^{-1} 
\ln \left(1 + Pe^2\tau \beta V_{\rm ext}(z)\right)\,. 
\end{eqnarray}
As for the effective interaction, eq.~\eqref{potential_pair}, the effective external 
field $V^{\rm eff}_{\rm ext}(z)$ vanishes wherever $V_{\rm ext}(z)$ is zero.

\subsection{Average interaction force} 
To approximate the interaction of ABPs in eq.~\eqref{average_interaction}, we make the \textit{adiabatic} assumption, which is standard within DDFT \cite{marconi1999,archer_evans_spinodal2004}, that
\begin{eqnarray}\label{ddft_approximation}
\F^{\rm int}(\rr,t)=\mu_{\rm ex}= -\nab\frac{\delta \mathcal{F}^{\rm eff}_{\rm ex}[\,\rho\,]}{\delta \rho(\rr,t)},
\end{eqnarray}
can be expressed as the gradient of a local equilibrium (excess) chemical potential  $\mu_{\rm ex}$.
We further assume that the excess Helmholtz free energy functional $\mathcal{F}^{\rm eff}_{\rm ex}$ it that of 
 particles interacting via the effective potential, eq.~\eqref{potential_pair}. 
Previous attempts have either used an adaptable parameter in $\F^{\rm int}(\rr,t)$ \cite{speck2014,bialke2013} 
or the excess free energy corresponding to the bare potential $u(r)$ \cite{pototsky2012}, which cannot account for MIPS.

Substituting eqs.~\eqref{onebody_active} and \eqref{ddft_approximation} into eq.~\eqref{DDFT_exact} yields 
\begin{eqnarray}\label{equation_of_motion}
\frac{\partial \rho(\rr,t)}{\partial t} = D_\text{t}\,\nab\cdot\left( \rho(\rr,t)
\nab\frac{\delta \beta\mathcal{F}^{\rm eff}[\,\rho\,]}{\delta \rho(\rr,t)}
\right).
\end{eqnarray}
The effective equilibrium free energy functional reads
\begin{eqnarray}\label{free_energy}
\mathcal{F}^{\rm eff}[\,\rho\,] = 
\mathcal{F}_{\rm id}[\,\rho\,] + \mathcal{F}_{\rm ex}^{\rm eff}[\,\rho\,] 
+\int\!\mathrm{d}\rr\, V_{\rm ext}^{\rm eff}(z)\rho(\rr)\,, 
\end{eqnarray}
where $\beta\mathcal{F}_{\rm id}[\,\rho\,] \!=\!\int \mathrm{d}\rr\, \rho(\rr)\left(\,  
\ln(\Lambda^3\rho(\rr))-1\right)$ 
is the exact ideal-gas contribution with the thermal wavelength $\Lambda$.

To arrive at a closed theory, we require an expression 
for $\mathcal{F}^{\rm eff}_{\rm ex}$ in eq.~\eqref{free_energy}
which describes the effective interaction and applies for all densities. 
In liquid-state theories, a potential of Lennard-Jones type, with its minimum at $r_0$, 
is usually treated by a separation, 
$u(r)\!=\!u_{\rm rep}(r)+u_{\rm att}(r)$, into repulsive and attractive contributions. 
We employ the Weeks-Chandler-Anderson prescription~\cite{WCA} by choosing 
$u_{\rm rep}(r)\!=\!u(r)\!-\!u(r_0)$ and $u_{\rm att}(r)\!=\!u(r_0)$ if $r\!<\!r_0$ and, 
otherwise, $u_{\rm rep}(r)\!=\!0$ and $u_{\rm att}(r)\!=\!u(r)$.
The soft repulsion is then mapped onto a system of hard spheres with suitably chosen 
diameter $\sigma$. 
This choice is not unique \cite{hansen_mcdonald1986}. 
We use the simple criterion, 
$\sigma\!=\!\int_0^{r_0} \mathrm{d}r\,(1-\exp(-\beta u_\text{rep}^{\rm eff}(r)))$,  
by Barker and Henderson \cite{barker_henderson1967}, 
which is both independent of the density and free of empirical parameters. 
The proposed separation yields the approximate functional 
\begin{eqnarray}
\!\!\!\!\!\mathcal{F}_{\rm ex}^{\rm eff}[\,\rho\,] = 
\mathcal{F}_{\rm ex}^{\rm (hs)}[\,\tilde{\rho}\,] 
+\! {\iint}\! \mathrm{d}\rr_1 \mathrm{d}\rr_2 \rho(\rr_1)\rho(\rr_2){\frac{u^{\rm eff}_{\rm att}(r_{12})}{2}}\,, \!\!\!\!\! \label{meanfield}
\end{eqnarray}
which provides a mean-field treatment of the attraction $u^{\rm eff}_{\rm att}(r_{12})$, where 
$r_{12}\!=\!|\rr_1-\rr_2|$. 
For the hard-sphere reference free energy, $\mathcal{F}_{\rm ex}^{\rm (hs)}[\,\tilde{\rho}\,]$, we employ 
the well-known Rosenfeld functional \cite{rosenfeld89}, 
where $\tilde{\rho}\!=\!\rho\, d^3\!/\sigma^3$.
With the form of eq.~\eqref{meanfield}, 
our aim is to provide a simple but qualitatively robust description of interacting ABPs (or OUPs).

\begin{figure}
\includegraphics[width=0.239\textwidth] {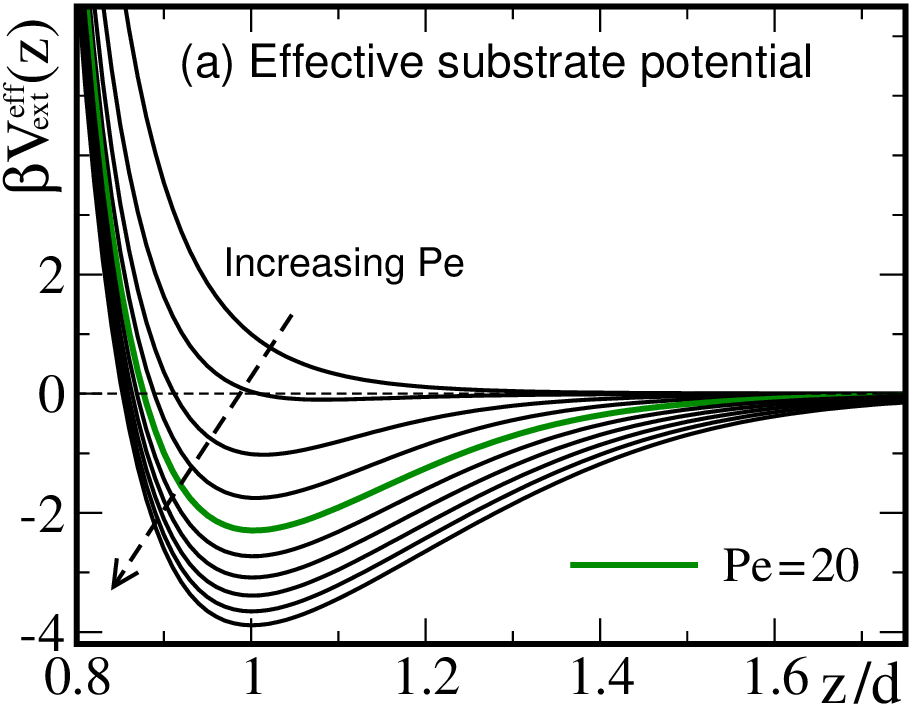} \hfill
\includegraphics[width=0.239\textwidth] {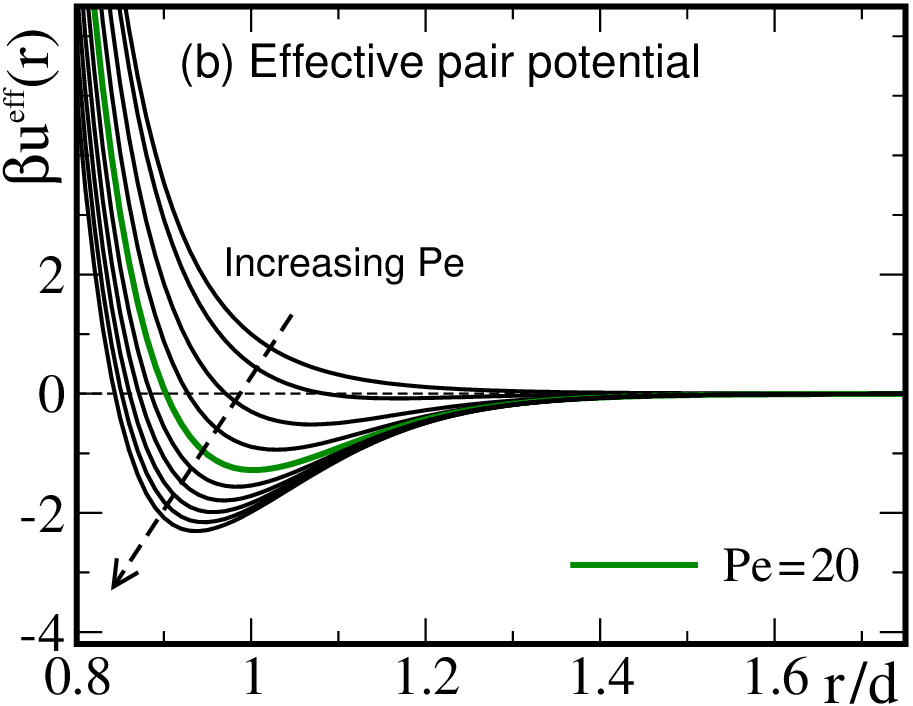} 


\caption{\label{fig_plots}(Colour online) Effective potentials of {\bf (a)} a single wall and {\bf (b)} between two particles,
given by eq.~\eqref{effective_external} and eq.~\eqref{potential_pair} with
$\beta V_{\rm ext}(z)\!=\!(z/d)^{-12}\!$ and
$\beta u(r)\!=\!(r/d)^{-12}$, respectively, $\tau\!=\!0.065$ and $Pe$ is increased from 0 to 45 in steps of 5.}
\end{figure}
\subsection{Effective equilibrium theory}

In the resulting equilibrium density functional theory (DFT) \cite{bob79,bob92} a minimisation $\delta\Omega[\rho]/\delta\rho(\rr)=0$ of the grand potential functional
$\Omega[\rho]\!=\!\mathcal{F}^{\rm eff}[\,\rho\,]\!-\!\int \!\mathrm{d}\rr \mu\rho(\rr)$ with $\mu$ being the chemical potential,
yields the (unique) inhomogeneous steady-state density profile.
A study of the dynamics via eq.~\eqref{equation_of_motion} is also possible, but goes beyond the 
scope of this work.
For our purposes we consider soft repulsive substrates.  
The one-body external field of a single wall is specified by 
$\beta V_{\rm ext}(z)\!=\!(z/d)^{-12}$ and that of a slit pore of width 
$W$ by $\beta V_{\rm ext}(z)\!=\!(z/d)^{-12}\!+\!((W\!-\!z)/d)^{-12}$.

We can rationalise some active phenomena alone from the effective potentials.
{According to fig.~\ref{fig_plots},} the evolution for a single wall is {qualitatively similar to that of the pair potential~\cite{faragebrader2015}:}
  as $Pe$ increases, $V^{\rm eff}_{\rm ext}(z)$ develops an attractive well. 
This reflects the higher probability of an ABP to be found in the vicinity of the 
wall~\cite{elgeti2013} 
{and the wall accumulation on the collective level~\cite{xiao2014,yang2014,ni2015}.}
It is also obvious from eq.~\eqref{effective_external} that we can qualitatively describe the experimental observation \cite{palacci2010,ginot2015} that activity counteracts the sedimentation in a gravitational field $\beta V_{\rm ext}(z)\!=\!mgz$.
The aim of {the present} study is, however, different.
In the following, we apply our DFT to interfaces in the vicinity of the binodal line, which requires the existence of MIPS in the bulk system.
Although being well understood in equilibrium, the resulting phenomena are not obvious from discussing the effective interactions alone.

\begin{figure}
\includegraphics[width=0.239\textwidth] {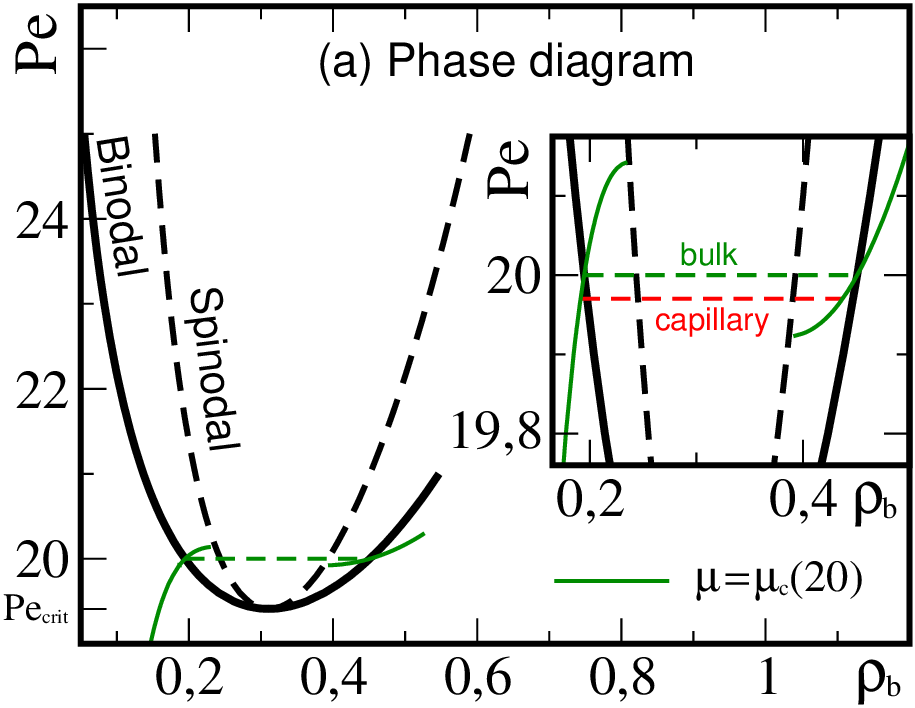} \hfill
\includegraphics[width=0.24\textwidth] {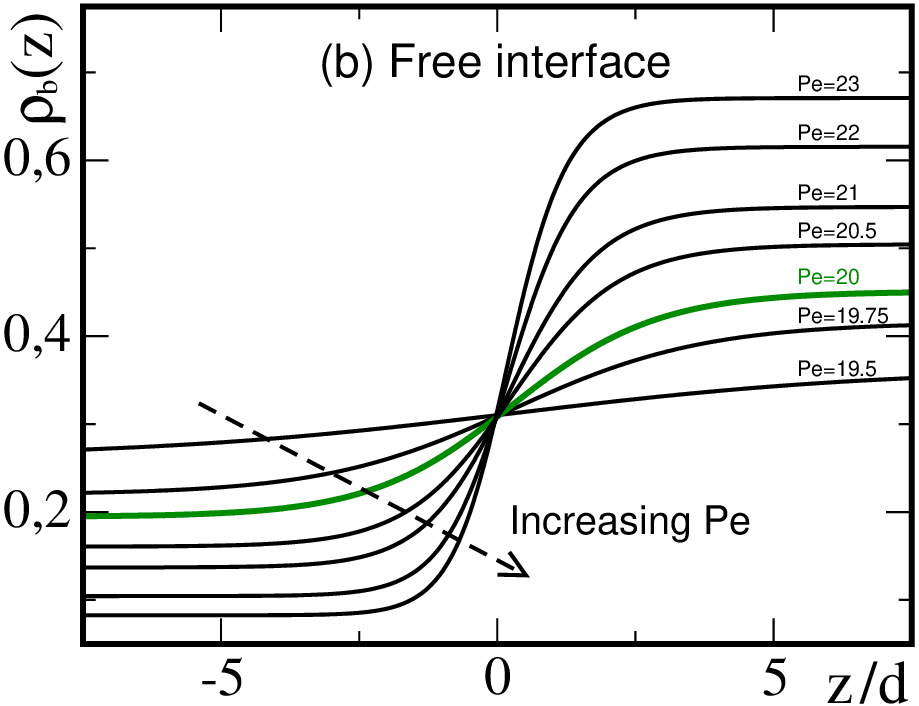} \\

\vspace*{-.25cm}

\includegraphics[width=0.24\textwidth] {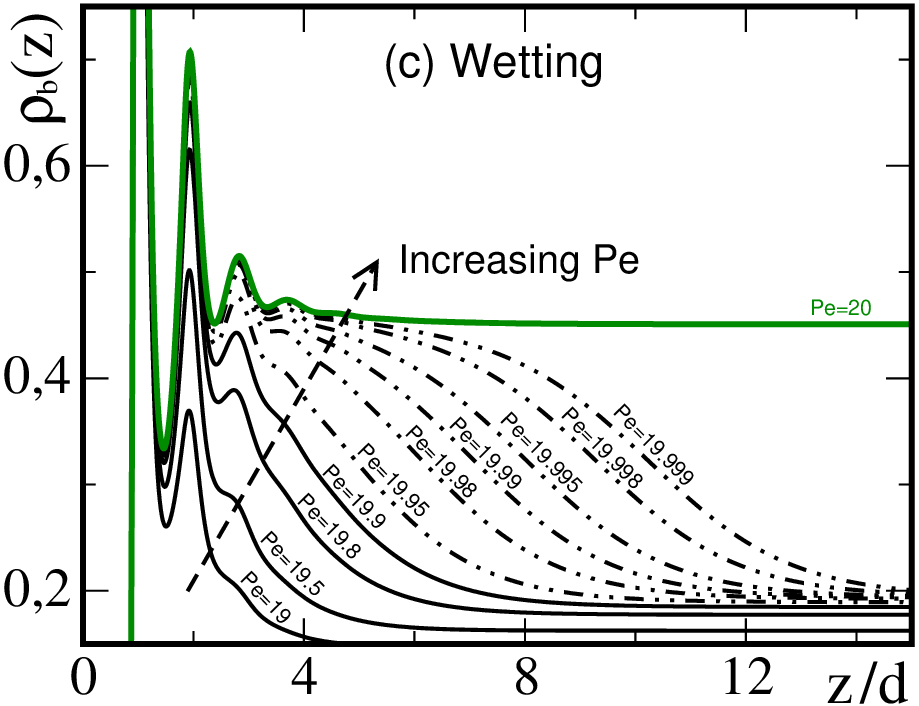} \hfill
\includegraphics[width=0.24\textwidth] {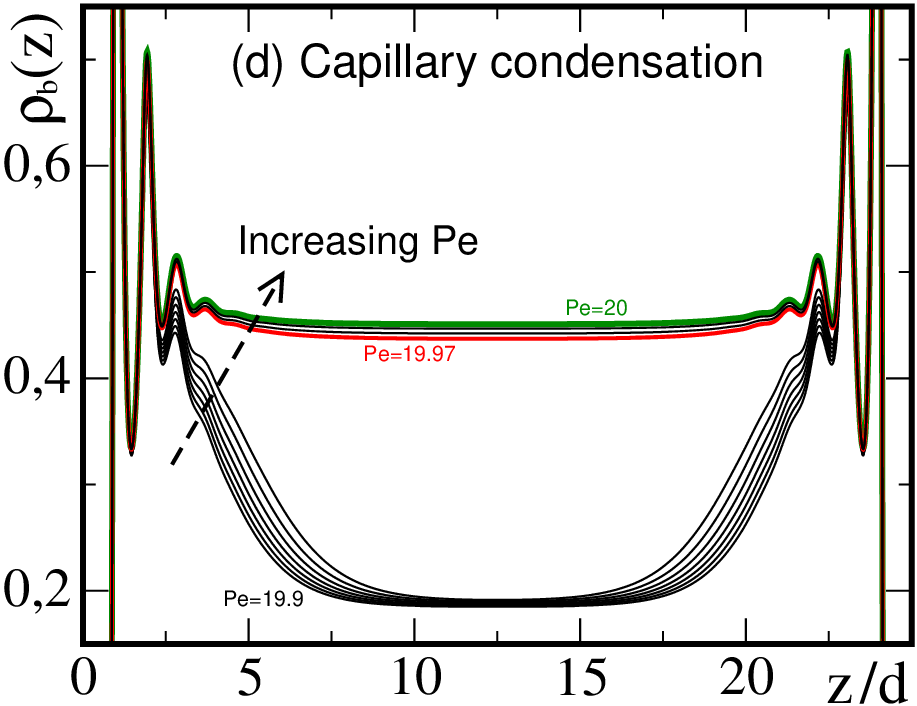}


\caption{\label{fig_z12}(Colour online) Phase behaviour for the effective potentials in fig.~\ref{fig_plots}.
{\bf (a)} Binodal and spinodal in the \mbox{$\rho_\text{b}$--$Pe$} plane ($\rho_\text{b}\!=\!\rho d^3$).
The thinner lines denote all (including metastable) points of equal chemical potential $\mu\!=\!\mu_\text{c}(20)$ touching the binodal at $Pe\!=\!20$. 
The closeup in the inset reveals the stable phases along this path in the bulk or in a capillary.
{\bf (b)} Free interfaces between colloid-rich and colloid-poor phases. 
{\bf (c)} Wetting near a single wall 
at $\mu\!=\!\mu_\text{c}(20)$. Dot-dot-dashed lines denote that the dense phase is metastable. 
{\bf (d)} Density in a slit (see text) of width $W\!=\!25d$ 
at $\mu\!=\!\mu_\text{c}(20)$, increasing $Pe$ from $19.9$ to $20$ (steps of $0.01$).
At $Pe\!=\!19.97$ (see fig.~\ref{fig_z12}a) capillary condensation occurs and the slit fills with the liquid phase. }
\end{figure}

\section{Passively repulsive system} 

We first consider passively repulsive spheres, $\beta u(r)\!=\!(r/d)^{-12}$, which 
exhibit MIPS in the bulk for $Pe\!>\!Pe_{\rm crit}\!=\!19.41$ within our DFT for a fixed $\tau\!=\!0.065$, see fig.~\ref{fig_z12}a.
Note that the critical point is located at a smaller $Pe$ than in ref.~\cite{faragebrader2015}, where 
a more accurate bulk integral-equation theory was employed.
However, the spinodals obtained in both approaches agree qualitatively.

\subsection{Free interface}

In fig.~\ref{fig_z12}b we show the density profiles at the free interface for different values of 
$Pe\!>\!Pe_{\rm crit}$. 
As could be anticipated from our effective thermodynamic description, 
the functional form is familiar from equilibrium studies and is consistent with recent simulations~\cite{speck_interface2014}. 
All plots in fig.~\ref{fig_z12}b are fitted well by a hyperbolic tangent profile, $\rho(z)\!\approx\!\rho_\text{g}\!+\!(\rho_\text{l}\!-\!\rho_\text{g}) 
(1\!+\!\tanh((z\!-\!\Delta z)/(2\delta))/2)$, where $\rho_\text{g}$ and $\rho_\text{l}$ are the 
coexisting (colloid-poor) gas and (colloid-rich) liquid densities, respectively, 
and $\delta$ is the interfacial width. 
The interface broadens as the value of $Pe$ is reduced, eventually diverging 
at the critical point. 
Within a mean-field description, the width diverges in the same fashion as the 
bulk correlation length \cite{bob92}, namely $\delta\!\sim\!(Pe\!-\!Pe_{\rm crit})^{-1/2}$.
Accordingly, the (thermodynamic) interfacial tension decreases to zero with the exponent $3/2$.

Simulations reveal strong interfacial fluctuations \cite{speck_interface2014},
which are omitted within mean-field theory.  
These long-wavelength fluctuations would lead to a broadening of the `intrinsic' 
interfacial profile generated by mean-field theory. 
 Although techniques exist to incorporate capillary waves \cite{schmidt_capillary2004}, 
it is not clear whether the interfacial fluctuations between the MIPS states are 
of capillary-wave type, or whether the interface is either {\it rough} or {\it smooth} \cite{bob92}.

\subsection{Wetting and capillary condensation}  
In fig.~\ref{fig_z12}c we show the steady-state density at a single wall as the binodal is approached on the path drawn in fig.~\ref{fig_z12}a, increasing 
$Pe$ at fixed $\mu$.
The strong first peak reflects the minimum in the effective wall potential in fig.~\ref{fig_plots}a.
For $Pe\!\gtrsim\!19.92$, where the liquid branch meets the spinodal, a macroscopic region of increased density 
emerges at the substrate, i.e., a wetting layer of the colloid-rich phase forms. 
Further approaching the binodal, the layer thickness grows significantly for $Pe\!\gtrsim\!19.98$ and eventually diverges at $Pe\!=\!20$, whereas the peaks near the wall practically remain identical.

In fig.~\ref{fig_z12}d we plot the density profiles under the same conditions
as in fig.~\ref{fig_z12}c, but now considering two repulsive walls at a separation 
$W\!=\!25d$. 
By increasing $Pe$ we observe the onset of wetting at both walls.
At $Pe\!\approx\!19.97$ the density profile jumps discontinuously and the slit fills with 
the colloid-rich phase, which is only metastable in the bulk (compare fig.~\ref{fig_z12}a): an activity-induced capillary condensation occurs. 
This discontinuous change in the density
leads to a jump in both excess adsorption and osmotic pressure acting on the walls.
The transition point shifts to even lower $Pe$ upon decreasing $W$ (not shown).

Wetting \cite{sullivan1979} and capillary condensation \cite{evans_tarazona1984} are quite familiar 
from equilibrium studies of attractive particles at attractive walls.
Here we emphasise that the behaviour is a consequence of the activity, represented by the effective interparticle and external potentials, and would be entirely absent in the purely repulsive passive system. 
Simulations reveal similar density profiles in \textit{two} dimensions, but
we do not reproduce the strong oscillations with $W$ in the surface tension resulting from crystalline order in the 
slit~\cite{ni2015}.

\begin{figure}
\includegraphics[width=0.24\textwidth] {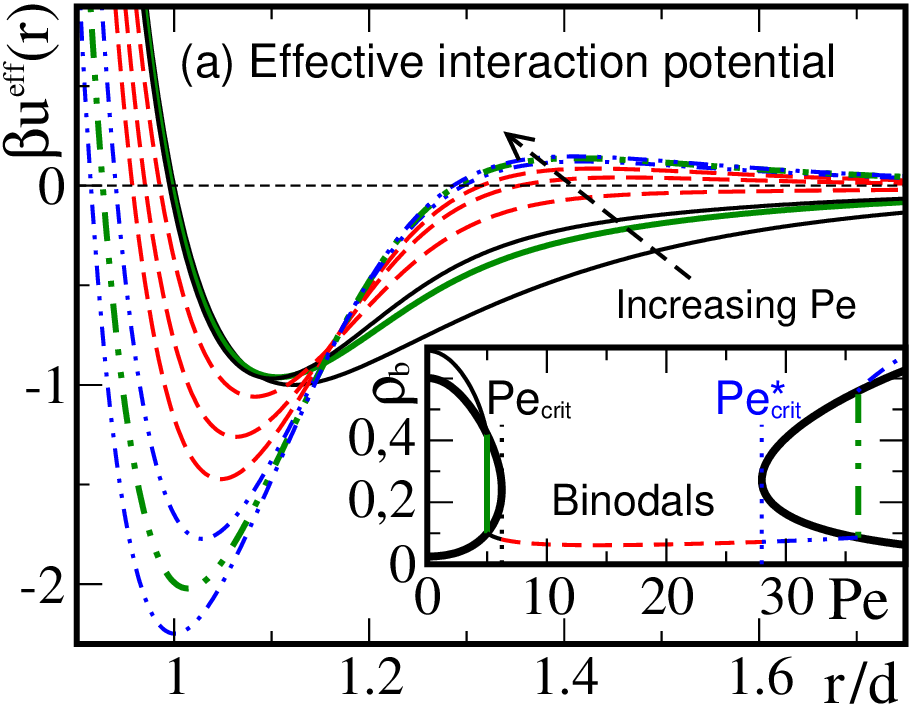} \hfill
\includegraphics[width=0.24\textwidth] {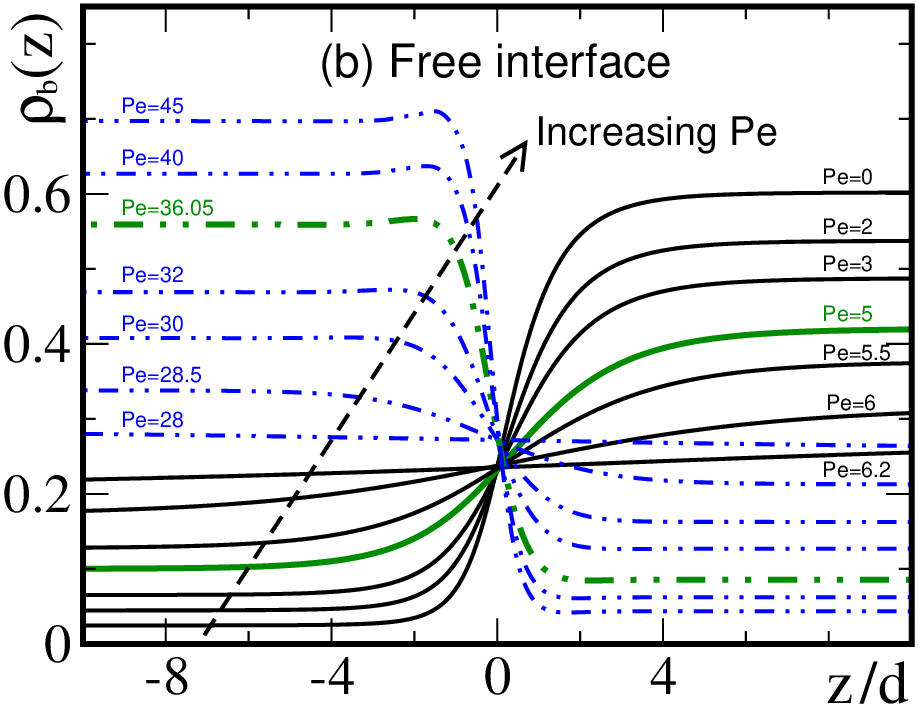} \\

\vspace*{-.25cm}

\includegraphics[width=0.24\textwidth] {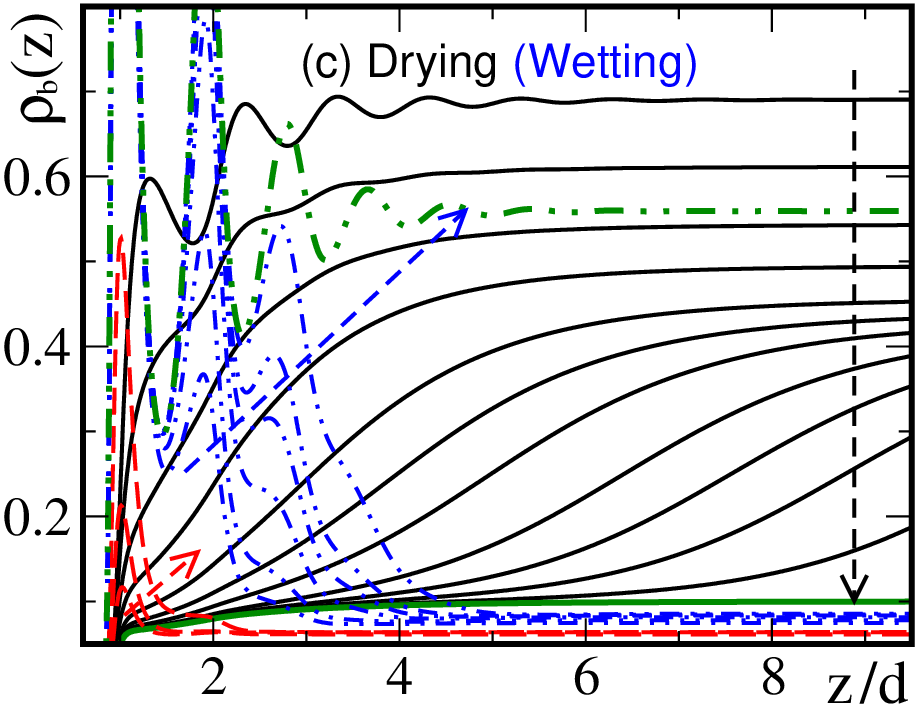} \hfill
\includegraphics[width=0.24\textwidth] {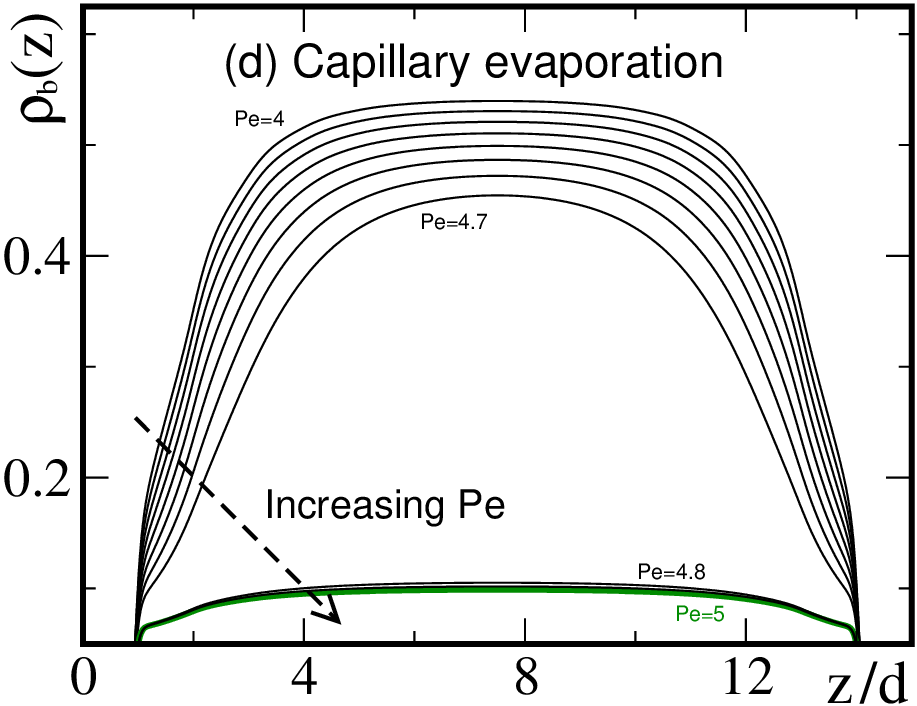} 


\caption{\label{fig_LJ} (Colour online) Density $\rho_\text{b}(z)\!=\!\rho(z)d^3$ for $\tau\!=\!0.045$ and different $Pe$. Pair potential $\beta u(r)\!=\!4\,(r/d)^{-12}-4\,(r/d)^{-6}$, external potential 
$\beta V_{\rm ext}(z)\!=\!(z/d)^{-12}\!$.  
{\bf (a)} Effective pair potential $u^{\rm eff}(r)$ 
for (i) $Pe\!\in\!\{0,5,6.2\}$, (ii) $Pe\!\in\!\{10,15,20\}$ and (iii) $Pe\!\in\!\{28,36.05,45\}$.
Inset: binodals in the \mbox{$Pe$--$\rho_\text{b}$} plane (thick curves) and points of equal chemical potential $\mu\!=\!\mu_\text{c}(5)$.
The (i) thin solid, (ii) dashed and (iii) dot-dot-dashed lines 
denote the regions bounded by the two critical points $Pe_\text{crit}$ and $Pe_\text{crit}^*$.
The thicker (straight) lines denote bulk coexistence at $Pe\!=\!5$ and $Pe\!=\!36.05$. 
{\bf (b)} Free interface for (i) $Pe\!\leq\!6.2$ (as in fig.~\ref{fig_z12}b) and (iii) $Pe\!\geq\!28$ (inverted at $z\!=\!0$).   
{\bf (c)} Density near a single wall at $\mu\!=\!\mu_\text{c}(5)$.
(i) Drying for $0\!\leq\! Pe\!\leq\!5$ (decreasing steps, compare fig.~\ref{fig_z12}c) (ii) increasing accumulation at the wall 
for $Pe\!\in\!\{10,15,20\}$ and (iii) wetting-like behaviour for $30\!\leq\! Pe\!\leq\!36.05$ 
(decreasing steps, compare fig.~\ref{fig_z12}c)
{\bf (d)} Density in a slit of width $W\!=\!15d$ at $\mu\!=\!\mu_\text{c}(5)$, increasing $Pe$ from $4$ to $5$ in steps of $0.1$.
At $Pe\!=\!4.74$ capillary evaporation occurs. }
\end{figure}

\section{Passively attractive system} 
 
 As shown in fig.~\ref{fig_LJ}a,
 the effective pair potential of an active Lennard-Jones fluid with $\beta u(r)\!=\!4\,(r/d)^{-12}-4\,(r/d)^{-6}$ and $\tau\!=\!0.045$ develops a repulsive tail for $Pe\!\gtrsim\!15$.
 The DFT from eq.~\eqref{meanfield} predicts an exotic phase diagram for such interactions~\cite{archer2007}. 
 We find two critical points in the active system at $Pe_\text{crit}\!=\!6.24$ and $Pe_\text{crit}^*\!=\!27.98$ accounting for a motility-induced suppression of the phase separation and a re-entrant MIPS, respectively.
In the latter case, the precise character of the transition is an open problem,
as the colloid-rich phase will very likely consist of clusters~\cite{faragebrader2015,archer2007}.

\subsection{ Free interface}

We first discuss fig.~\ref{fig_LJ}b for the free interface, which also reflects the 
rich bulk phase behaviour from fig.~\ref{fig_z12}a.
The shape of the density profiles at small $Pe$ is similar to that in the passively repulsive case (fig.~\ref{fig_z12}b) although the response to a change in activity is the opposite.
At $Pe_\text{crit}\!<\!Pe\!<\!Pe_\text{crit}^*$, there is no phase separation, until a new interface develops at higher activity.
The density profile of this re-entrant interface is non-monotonic, which is most pronounced on the 
colloid-rich side.
Moreover, the bulk correlation length, i.e., the interface width, is markedly smaller than at 
lower activity,  
which we see when comparing characteristic profiles.
This could already be anticipated from the smaller effective diameter $\sigma$ of the reference fluid 
at higher activity (compare fig.~\ref{fig_LJ}a).

 The interfacial tension is smaller at higher activity, consistent with the fact 
that activity consumes energy from the system.
To substantiate this observation we study the near-critical behaviour.
Approaching $Pe_\text{crit}$, the interfacial tension $\gamma$ is about 20 times higher than for $Pe_\text{crit}^*$, which we see by comparing the prefactors $\gamma_0$ and $\gamma_0^*$ in the fits to $\gamma\!=\!\gamma_0\,(Pe_{\rm crit}\!-\!Pe)^{3/2}$ and $\gamma\!=\!\gamma_0^*\,(Pe\!-\!Pe_{\rm crit}^*)^{3/2}$ for 
sufficiently close sets of data points.

\subsection{ Drying and capillary evaporation} 

In the following, we fix $\mu$ to find phase coexistence at $Pe\!=\!5$ and $Pe\!\approx\!36.05$.
Figure~\ref{fig_LJ}c illustrates the wide variety of phenomena predicted for a passively attractive 
fluid at a passively repulsive wall when following the path depicted in fig.~\ref{fig_LJ}a. 
At zero activity, $Pe\!=\!0$, the density of a pure Lennard-Jones fluid is depleted close to the wall, 
as the attractive interparticle interaction favours particle cohesion.
For higher $Pe$, the bulk density decreases and 
 a region of the colloid-poor phase begins to develop, as it becomes metastable in the bulk for $Pe\!\gtrsim\!4.5$, ultimately diverging
 at $Pe\!=\!5$.
As in fig.~\ref{fig_plots}a, the effective substrate potential is barely attractive for 
$Pe\!\lesssim\!5$. 
Hence, the particles still prefer to accumulate next to each other.
The first main peak in fig.~\ref{fig_LJ}c has developed for $Pe\!=\!10$.
Increasing $Pe$ it becomes higher, reflecting the stronger 
effective attraction of the wall.

 Beyond $Pe_\text{crit}^*$, we also observe an increase in height of the second and third peaks 
before an infinite layer of the colloid-rich phase appears at $Pe\!\approx\!36.05$. 
In contrast to fig.~\ref{fig_z12}c, this wetting-like behaviour occurs abruptly
due to the highly attractive effective wall potential and the relatively small bulk 
correlation length (deduced from fig.~\ref{fig_LJ}b).
Thus a wetting film would be merely thin compared to the drying film for $Pe\!<\!5$.
In any case, arguing about a \textit{true} re-entrant wetting phenomenon at $Pe\!>\!28$ remains rather 
speculative, due to the possible influence of clustering~\cite{archer2007}.

Finally, we predict that a passively attractive system (in a liquid-like bulk state) confined between two parallel plates evaporates to a colloid-poor phase at a certain value of $Pe<Pe_\text{crit}$ (see fig.~\ref{fig_LJ}d), even before the activity suppresses the phase separation in the bulk.
For the chosen example, the capillary evaporation in a slit of width $W\!=\!15d$ occurs at 
$Pe\!\approx\!4.74$, whereas the line in fig.~\ref{fig_LJ}a meets the binodal only at $Pe\!=\!5$.
Given the results shown in fig.~\ref{fig_LJ}c one can also envisage a re-entrant 
capillary-condensation transition (not shown in fig.~\ref{fig_LJ}d).

\section{Discussion} 

 Our focus so far has been on the density distribution of ABPs,
calculated with an effective thermodynamic theory.
The definition of pressure~\cite{cates_kardar,takatori2014,takatori2015,solon_BrownianPressure2015} and interfacial 
tension~\cite{speck_interface2014} within such a framework remains the subject of debate~\cite{catesarxiv,speckarxiv}. 
In the following, we argue in how far our theory captures the physics of ABPs and how it connects to some more direct approaches.

\subsection{Effective equilibrium regime}

 The bulk theory addresses coupled OUPs. Although the exact configurational probability distribution does not satisfy a Fokker-Planck equation, the 
optimal Markovian (Fox) approximation~\cite{faragebrader2015} does so and thus 
provides the possibility of an equilibrium description~\cite{catesarxiv}.
From eq.~\eqref{smol_eq2} we have derived an effective external field to complement our bulk theory.
 Despite our approximations and certain discrepancies between the steady-states of ABPs~\cite{solonEPJST} and OUPs~\cite{szamel2014},
we are confident that our {predictions} in figs.~\ref{fig_z12} and~\ref{fig_LJ} of the non-equilibrium
collective 
behaviour {are} qualitatively robust for both types of self-propulsion mechanisms, as we assume  moderate deviations from equilibrium {and simulations in two dimensions indicate a similar phenomenology~\cite{ni2015,yang2014}.}

 We note that one prediction of the bulk theory omitted so far is an effective diffusion coefficient~\cite{faragebrader2015},
generalising $\mathcal{D}(r)$ entering in eq.~\eqref{potential_pair}.
 It allows us to distinguish between the
ideal Brownian $\beta p_\text{B}\!=\!\rho$ and
ideal swim pressure $\beta p_\text{S}^\text{(id)}\!=\!\rho Pe^2\tau/3$~\cite{marconi,takatori2014,takatori2015,solon_BrownianPressure2015,speck_interface2014}
by setting the energy scale via the effective temperature $\beta_\text{eff}^{-1}=\beta^{-1}(1\!+\!Pe^2\tau/3)$, 
 also employed~\cite{marconi,ginot2015,palacci2010,szamel2014} to describe active sedimentation in certain limits~\cite{enculescu2011,solonEPJST}. 
An effective external field for OUPs, consistent with ref.~\cite{faragebrader2015}, 
and a more general diffusion coefficient are addressed in Refs.~\cite{EEA1,EEA2}.

Previous works (see, e.g.,~refs.~\cite{cates_tailleur2014,cates2013,solonEPJST}) have proposed an exact mapping to a \textit{local} 
bulk free energy functional by assuming a slowing down of the particles due to collisions.
In contrast, we employ an effective potential in eq.~\eqref{ddft_approximation}.
We stress that our free energy, eq.~\eqref{meanfield}, is built on weighted densities, 
i.e., convolutions of the density with geometrical measures of a sphere~\cite{rosenfeld89}, and is thus intrinsically non-local.
Hence, it treats homogeneous and inhomogeneous situations on an equal footing, avoiding the need to add square-gradient terms 
\cite{cates_tailleur2014,solonEPJST} to a local theory.
The free energy in eq.~\eqref{free_energy} combines the effective forces derived from independent approximations
for bulk interactions and external fields. 
It thus approximately compensates the information lost when integrating out the orientational degrees of freedom, 
e.g., the anisotropy near boundaries~\cite{elgeti2013}.

\begin{figure}
\includegraphics[height=0.13\textwidth]{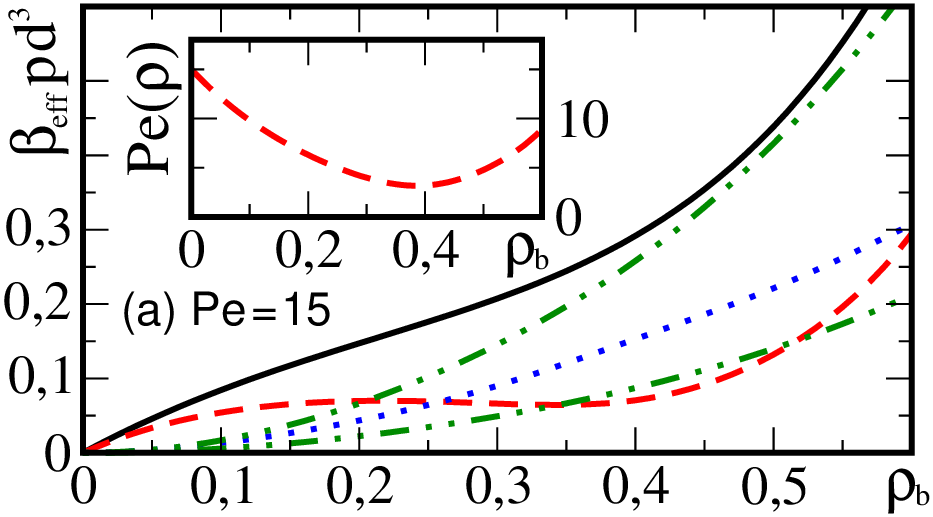} \hfill \includegraphics[height=0.13\textwidth]{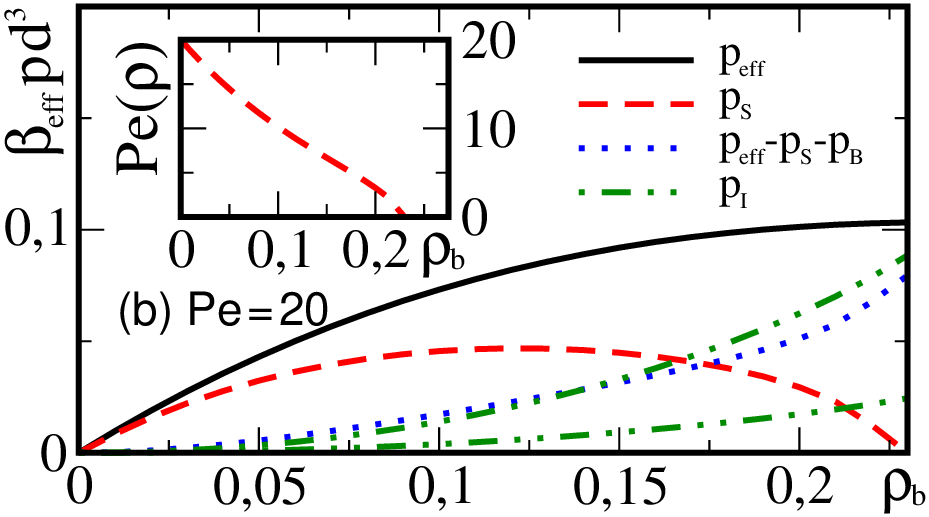} 
\caption{\label{fig_Pressure} 
 (Colour online) Effective bulk pressure $p_\text{eff}$ compared to swim pressure $p_\text{S}$, interaction pressure $p_\text{I}$ 
 {(small values for hard-sphere potential)}
 and ideal Brownian \mbox{pressure $p_\text{B}$ (see text) using} 
 $\beta u(r)\!=\!(r/d)^{-12}$, $\tau\!=\!0.065$ and {\bf (a)} $Pe\!=\!15$ or {\bf (b)} $Pe\!=\!20$.
The insets show the projected velocity $Pe(\rho)$ (see text).}
\end{figure}

\subsection{Swim pressure}

In dense systems, the interplay between the bare interaction pressure $p_\text{I}$ 
and $p_\text{S}$ 
has been proposed as the driving force of MIPS~\cite{takatori2014,solon_BrownianPressure2015}. 
 As the swim pressure arises from the coupling between particle position and orientation~\cite{takatori2014,speck_interface2014}, 
 the latter being not resolved in our theory~\cite{faragebrader2015},
it is not directly accessible. 
  However, $\beta p_\text{S}\!=\!\rho Pe(\rho)Pe\,\tau/3$ can be expressed~\cite{solon_BrownianPressure2015} 
  in terms of a density-dependent projected velocity $Pe(\rho)$~\cite{stenhammar2014,cates_tailleur2014}.
  Comparing eq.~\eqref{ddft_approximation} to refs.~\cite{cates_tailleur2014,cates2013} \mbox{(in our case $\mu_{\rm ex}\!=\!0$ if $\rho\!=\!0$)} yields $Pe^2(\rho)\!=\!(Pe^2\!+\!3/\tau)\exp(2\beta\mu_{\rm ex})\!-\!3/\tau$.
  The virial pressure {$\beta p_\text{I}\!=\!(2\rho^2\pi/3) \int\!\upd r\,r^3g(r)\,\partial_ru(r)$~\cite{hansen_mcdonald1986}
    with} the effective radial distribution $g(r)$,
    obtained with DFT methods~\cite{bob79,bob92} and the Ornstein-Zernicke equation~\cite{hansen_mcdonald1986} {(we set $g(r\!<\!\sigma)\!=\!0$),
    is calculated either for the bare potential $u(r)$ or, more consistently with our DFT implementation, for hard spheres of} diameter $\sigma(Pe\!=\!0)$. 
{Note that $g(r)$ and thus $p_\text{I}$ is ill-defined within the spinodal.}

Regarding the shortcomings of mean-field theory, {underestimating $\mu_\text{ex}$ and leaving 
it unclear how to ideally calculate $p_\text{I}$,} 
fig.~\ref{fig_Pressure} certifies a reasonable agreement of $p_\text{I}\!\simeq\!p_\text{eff}\!-\!p_\text{S}\!-\!p_\text{B}$,
with the overall effective pressure $p_\text{eff}$, minimising our functional on the active energy scale $\beta_\text{eff}^{-1}$.
The {swim pressure $p_\text{S}(\rho)$ behaves reasonably~\cite{takatori2014,solon_BrownianPressure2015} up to the minimum of the velocity $Pe(\rho)$, which, above a certain $Pe\!>\!Pe_\text{crit}$}, decreases almost linearly to zero~\cite{cates_tailleur2014,fily2012,cates2013,stenhammar2014}.
A similar procedure for inhomogeneous situations, {while implementing the aforementioned improvements, 
could clarify in how far our free-energy based theory consistently describes
the mechanical pressure exerted on a wall~\cite{cates_kardar}
and the interfacial tension \cite{speck_interface2014,speckarxiv} in active systems.
Some approximate expressions for these quantities were recently derived~\cite{marconi2016}
from another effective treatment of the OUPs model~\cite{maggi,marconi}
without explicitly employing effective pair interactions.
}

\section{ Conclusions}

Using an effective equilibrium description
we predict that ABPs (or other systems showing bulk MIPS) will exhibit a variety of interfacial phase transitions  induced by activity. 
To construct a DDFT for inhomogeneous active systems, 
we derive from the position and orientation-resolved Fokker-Planck equation 
an effective external potential~\cite{pototsky2012}, which we 
combine with the effective interparticle potential from ref.~\cite{faragebrader2015}.
We discussed some caveats and {quantitative issues} but we are not aware of any alternative first-principles theory capable of dealing with 
MIPS either in the bulk or at interfaces. 

The tendency of repulsive particles to accumulate even at repulsive substrates for finite activity
can be seen directly from our density profiles and understood 
intuitively from comparing the effective 
potentials, as shown in fig.~\ref{fig_plots}.
We further anticipate that active fluids exhibit motility-induced wetting phenomena 
and, in a slit pore, capillary condensation. 
It would also be interesting to investigate the capillary crystallisation observed for 
two-dimensional ABPs \cite{ni2015}.
 At a sufficiently low P\'eclet number, the accumulation 
 near the wall can be suppressed when 
 the bare interaction between the particles is already attractive.
An experimental system similar to that used in ref.~\cite{schwarz-linek2012} may well show drying and 
capillary evaporation, when one or two opposing repulsive planar substrates are introduced.

As a next step we will {develop new grand-canonical-type simulation methods in three dimensions to}
test our predictions.
A natural continuation of the present work would be to investigate dynamical properties, although it is 
unclear to what extent the effective equilibrium picture remains valid beyond steady states. Moreover, 
it would be interesting to extend our current methods to study mixtures and to exploit recent work 
in equilibrium DFT~\cite{wittmann16} for treating anisotropic swimmers~\cite{ramaswamy2010,swim_aniso}.


\end{document}